\documentclass[graybox]{svmult}

\usepackage{mathptmx}       
\usepackage{helvet}         
\usepackage{courier}        
\usepackage{setspace}

\usepackage{makeidx}         
\usepackage{graphicx}        
\usepackage{multicol}        
\usepackage[bottom]{footmisc}

\usepackage{amsmath}
\usepackage{amssymb}

\makeindex             

\begin{document}

\title*{Elastic Network Models: Theoretical and Empirical Foundations}
\titlerunning{Elastic Network Models} 
\author{Yves-Henri Sanejouand}
\authorrunning{Y.-H. Sanejouand} 
\institute{Yves-Henri Sanejouand \at CNRS-UMR 6204, Facult\'e des Sciences, Nantes. \email{yves-henri.sanejouand@univ-nantes.fr}}
%
%
\maketitle

\abstract{Fifteen years ago Monique Tirion showed that the low-frequency normal modes of a protein are not significantly altered when non-bonded interactions are replaced by Hookean springs, for all atom pairs whose distance is smaller than a given cutoff value. Since then, it has been shown that coarse-grained versions of Tirion's model are able to provide fair insights on many dynamical properties of biological macromolecules. In this chapter, theoretical tools required for studying these so-called Elastic Network Models are described, focusing on practical issues and, in particular, on possible artifacts. Then, an overview of some typical results that have been obtained by studying such models is given.}


\section{Introduction}
\label{sec:intro}

In 1996, Monique Tirion showed that the low-frequency normal modes of a protein (see section \ref{ssec:nma}) are not significantly altered when Lennard-Jones and electrostatic interactions are replaced by Hookean (harmonic) springs, for all atom pairs whose distance is smaller than a given cutoff value~\cite{Tirion:96}. In the case of biological macromolecules, this seminal work happened to be the first study of an Elastic Network Model (ENM). The ENM considered was an all-atom one, chemical bonds and angles being kept fixed through the use of internal coordinates, as often done in previous standard normal mode studies of proteins~\cite{Go:82,Go:83,Levitt:83}. 

Soon afterwards, several coarse-grained versions of Tirion's ENM were proposed, in which each protein amino-acid residue is usually represented as a single bead and where most, if not all, chemical "details" are disregarded~\cite{Bahar:97,Hinsen:98}, including atom types and amino-acid masses. 

Since then, it has been shown that such highly simplified protein models are able to provide fair insights on the dynamical properties of biological macromolecules~\cite{Bahar:97,Maritan:02,Phillips:07,NMA}, including those involved in their largest amplitude functional motions~\cite{Tama:01,Gerstein:02}, even in the case of large assemblies like RNA polymerase II~\cite{Delarue:02}, transmembrane channels~\cite{Helene:03,Elnemo2}, whole virus capsids~\cite{Tama:02v} or even the ribosome~\cite{Tama:03}. As a consequence, numerous applications have been proposed, noteworthy for exploiting fiber diffraction data~\cite{Tirion:95}, solving difficult molecular replacement problems~\cite{Elnemo1,Delarue:04}, or for fitting atomic structures into low-resolution electron density maps~\cite{Delarue:04,Tama:04,Hinsen:05, Tama:05, Norma}.

However, the idea that simple models can prove enough for capturing major properties of objects as complex as proteins had been put forward well before Tirion's introduction of ENMs in the realm of molecular biophysics. In the following, after a brief account of previous results supporting this claim (section \ref{sec:back}), theoretical tools required for studying an ENM are described (section \ref{sec:meth}), focusing on practical issues and, in particular, on possible artifacts. Then, an overview of typical results that have been obtained by studying protein ENMs is given (section \ref{sec:emp}).

\section{Background}
\label{sec:back}

Indeed, coarse-grained models of proteins had been considered twenty years before M. Tirion's work, for studying what may well be the most complex phenomenon known at the molecular scale, namely, protein folding. Indeed, as soon as 1975, Michael Levitt and Arieh Warshel proposed to model a protein as a chain of beads, each bead corresponding to the C$_{\alpha}$ atom of an amino-acid residue, the centroid of each amino-acid sidechain being taken into account with another bead grafted onto the chain~\cite{Levitt:75}. That same year, Nobuhiro Go and his collaborators proposed an even simpler model in which the chain of beads is mounted on a two-dimensional lattice, each bead corresponding either to a single residue or, more likely, to a secondary structure element (e.g., an $\alpha$-helix) of a protein~\cite{Go:75}. Moreover, while the Levitt-Warshel model had been designed so as to study a specific protein, that is, a polypeptidic chain with a given sequence of amino-acid residues,  the Go model focuses on the conformation of the chain, more precisely, on the set of pairs of amino-acids that are interacting together in the chosen (native) structure. 

\begin{svgraybox}
So, it is fair to view protein ENMs as off-lattice versions of the Go model. 
\end{svgraybox}

Lattice models of proteins have been studied extensively since then so as to gain, for instance, a better understanding of the sequence-structure relationship. Noteworthy, if the chain is short enough, all possible conformations on the lattice can be enumerated, allowing for accurate calculations of thermodynamic quantities and univoqual determination of the free energy minimum. Moreover, if the number of different amino-acids is small enough, then the whole sequence space can also be addressed. For instance, in the case of the tridimensional cubic lattice, a 27-mer chain has 103346 self-avoiding compact (i.e. cubic) conformations~\cite{Gutin:90}. On the other hand,  if only two kinds of amino-acids are retained, that is, if only their hydrophobic or hydrophilic nature is assumed to be relevant for the understanding of protein stability, then a 27-mer has $2^{27}$ different possible sequences. This is a large number, but it remains small enough so that for each sequence the lowest-energy compact conformation can be determined and, when a nearly-additive interaction energy is considered~\cite{Wingreen:97}, the conclusion of such a systematic study happens to be an amazing one. Indeed, it was found that a few conformations (1\% of them) are "preferred" by large sets of sequences~\cite{Wingreen:96}. Moreover, although each of these sets forms a neutral net in the sequence space, it is often possible to "jump" from a preferred conformation to another, as a consequence of single-point mutations~\cite{Trinquier:99}. 

While the former property is indeed expected to be a protein-like one, allowing to understand why proteins are able to accomodate so many different single-point mutations without significant loss of both their structure and function, it is only during the last few years that the latter one has been exhibited. In particular, using sequence design techniques, a pair of proteins with 95\% sequence identity, but different folds and functions, was recently obtained~\cite{Orban:08}. 
If generic enough, such a property would help to understand how the various protein folds nowadays found on earth may have been "discovered" during the earliest phases of life evolution (e.g. prebiotic ones), since discovering a first fold could have proved enough for having access to many other ones, a single-point mutation after another. 

\begin{svgraybox}
In any case, this example shows how the study of simple models can help to think about, and maybe to understand better, major protein properties, in particular because such models can be studied on a much larger scale than actual proteins.
\end{svgraybox}

\section{Theoretical foundations}
\label{sec:meth}

The vast majority of protein ENM studies rely on Normal Mode Analysis (NMA)~\cite{NMA}. Moreover, the hypotheses underlying this kind of analysis probably inspired the design of the first ENM. Actually, in her seminal work, M. Tirion performed NMA in order to show that similar results can be obtained by studying an ENM or a protein described at a standard, semi-empirical, level~\cite{Tirion:96}. 
So, hereafter, the principles of NMA are briefly recalled (more details can be found in classic textbooks~\cite{Goldstein:50,Wilson:55}). Next, the close relationship between NMA and the different types of ENMs is underlined.

\subsection{Normal Mode Analysis}
\label{ssec:nma}

Newton's equations of motion for a set of $N$ atoms can not be solved analytically when $N$ is large (namely, $ N > 2$),
except in rare instances like the following, rather general, one. Indeed, for small enough displacements of the atoms in the vicinity
of their equilibrium positions, $V$, the potential energy of the studied system, can be approximated by the first terms of a Taylor series: 
\begin{equation}
\label{taylor}
V = V_0 + \sum_{i=1}^{3N} \left(\frac{\partial V}{\partial r_i}\right)_0 (r_i - r_i^0) +
  \frac{1}{2} \sum_{i=1}^{3N} \sum_{j=1}^{3N} \left(\frac{\partial^2 V}
  {\partial r_i \partial r_j}\right)_0 (r_i - r_i^0) (r_j - r_j^0) 
\end{equation}
where $r_i$ is the $i^{th}$ coordinate,  $r_i^0$, its equilibrium value, and $V_0$, 
the potential energy of the system at equilibrium.  

Since, within the frame of classical physics, the exact value of $V$ is meaningless (only potential energy differences are expected to play a physical role),
$V_0$ can be zeroed.  Moreover, since $V_0$ is a minimum of $V$, for each coordinate:
\begin{equation}
\notag
\left(\frac{\partial V}{\partial r_i}\right)_0 = 0
\end{equation}
This yields:
\begin{equation}
\label{quad}
V = \frac{1}{2} \sum_{i=1}^{3N} \sum_{j=1}^{3N} \left(\frac{\partial^2 V}
  {\partial r_i \partial r_j}\right)_0 (r_i - r_i^0) (r_j - r_j^0) 
\end{equation}

\noindent
In other words, if the atomic displacements around an equilibrium configuration are small enough,
then the potential energy of a system can be approximated by a quadratic form.

On the other hand, 
if the system is {\it not} under any constraint with an {\it explicit} time-dependence, then 
its kinetic energy can also be written as a quadratic form~\cite{Goldstein:50} and
it is straightforward to show that, when both potential and kinetic energy functions are quadratic forms, then
the equations of atomic motion have the following, analytical, solutions~\cite{Goldstein:50,Wilson:55,Sanejouand:Phd}:
\begin{equation}
\label{nma:solution}
r_i(t) = r_i^0 + \frac{1}{\sqrt{m_i}}
       \sum_{k=1}^{3N} C_k a_{ik} cos( 2 \pi \nu_k t + \Phi_k )
\end{equation}
where $m_i$ is the atomic mass and where $C_k$ and $\Phi_k$, the amplitude and phasis of the so-called normal mode of vibration $k$,
depend upon the initial conditions, that is, upon atomic positions and velocities at time $t=0$. Noteworthy,
$C_k$ is a simple function of $E_k$, the total energy of mode $k$. In particular, if all modes have identical total energies,
then: 
\begin{equation}
\label{qampl}
C_k = \frac{\sqrt{2 k_B T}}{2 \pi \nu_k}
\end{equation}
where $T$ is the temperature and $k_B$ the Boltzmann constant.
This means that the amplitude of mode $k$ goes as the inverse of its frequency, $\nu_k$.
As a matter of fact, when NMA is performed in the case of proteins, using standard all-atom force-fields,  
it can be shown that modes with frequencies below 30-100 cm$^{-1}$ are responsible  
for 90-95\% of the atomic displacements~\cite{Levy:82}. 

\begin{svgraybox}
Note that such analytical solutions can provide various
thermodynamic quantities like entropy, enthalpy, etc, and this, even at a quantum mechanical level of description~\cite{Levy:82}.
\end{svgraybox}

In practice, the $a_{ik}$'s involved in eq.~\ref{nma:solution}, which give the coordinate contributions to mode $k$, are 
obtained as the $k^{th}$ eigenvector of ${\bf H}$, the mass-weighted Hessian of the potential energy,
that is, the matrix whose element $i j$ is:
\begin{equation}
\label{secder}
 \left(\frac{\partial^2 V} {\sqrt{m_i m_j} \partial r_i \partial r_j}\right)_0 
\end{equation}
By definition, the $3N$ eigenvectors of a matrix like ${\bf H}$ form an orthogonal basis set. This means that, when $k \ne l$: 
\begin{equation}
\notag
 \left(\frac{\partial^2 V} {\partial q_k \partial q_l}\right)_0 = 0
\end{equation}
where $q_k$ is the so-called normal coordinate, obtained by projecting the $3N$ mass-weighted cartesian coordinates onto eigenvector $k$, namely:
\begin{equation}
\label{qr}
q_k = \sum_i^{3N} a_{ik} \sqrt{m_i} ( r_i - r_i^0 )
\end{equation}

Moreover, the eigenvalues of ${\bf H}$, that is, 
the diagonal elements of the matrix obtained by expressing ${\bf H}$ in this new basis set,
provide the $3N$ frequencies of the system since, for each mode $k$: 
\begin{equation}
\notag
 \left(\frac{\partial^2 V} {\partial^2 q_k}\right)_0 = (2 \pi \nu_k)^2
\end{equation}

The eigenvalues and eigenvectors of a matrix are obtained by an operation called a diagonalization.
In principle, for a real and symmetrical matrix like ${\bf H}$, such an operation is always possible. 
At a practical level, when the matrix size is not too large, that is, if the matrix can be stored in the computer memory,
algorithms and methods available in standard mathematical packages allow to get its eigenvalues and eigenvectors
at a CPU cost raising as $n N^2$, where $n$ is the number of requested eigensolutions.
In other words,  it is rather straightforward to obtain analytical solutions for the atomic motions, as long as
small-amplitude displacements around a given, well-defined,
equilibrium configuration are considered. 
Note that for a tridimensional system at equilibrium, at least six zero eigenvalues have to be obtained
(except if the system is linear, in which case there are five of them), corresponding to the six possible rigid-body motions  (translations or rotations)
of the entire system. However, if the system is {\it not} at equilibrium, negative eigenvalues are usually observed. Moreover, significant mixing
between rotation modes and some others can occur, leaving three zero eigenvalues only, that is, those corresponding to the
three translation modes of the system~\cite{Sanejouand:Phd}.

\begin{svgraybox}
The main drawback of NMA is obvious: the actual dynamics of a protein is much more complicated than assumed above.
As a matter of fact, even on the short timescales considered within the frame of standard molecular dynamics simulations,
a protein is able to jump from the attraction basin of an equilibrium configuration to another~\cite{Elber:87}, and the number of these
equilibrium configurations is so huge that it is unlikely for a nanosecond trajectory to visit one of them twice.
In other words, while NMA focuses on protein dynamics at the level of a single minimum of the potential energy surface (PES),
it is well known that for proteins at room temperature the relevant PES is a higly complex, multi-minima, one. 
\end{svgraybox}

NMA has several other drawbacks. 
For instance, starting from a given protein structure, e.g., as found in the Protein Databank (PDB), an equilibrium configuration
has to be reached. This is usually done using energy-minimization techniques. As a consequence, the structure
studied with NMA and a standard force field is always a distorted one, the C$_{\alpha}$ root-mean-square deviation (C$_{\alpha}$-r.m.s.d) from the initial structure  
being typically of 1-2{\AA}~\cite{NMA}.  

More importantly, within the frame of NMA, it is not obvious to take solvent effects into account, as the meaning
of an equilibrium configuration in the case of an ensemble of molecules in the liquid state is unclear. As a matter of fact, the first
NMA studies of proteins were performed {\it in vacuo}~\cite{Go:82,Go:83,Levitt:83,Brooks:83}. Note that, nowadays, the availability of implicit solvent
models, like EEF1~\cite{EEF1}, offers a more satisfactory alternative.   

However, as shown below, the main idea underlying the design of protein ENMs is not only to ignore the well-known drawbacks of NMA but, building upon its empirical successes, to add a few more on top of them.  

\subsection{The Elastic Network Model}
\label{ssec:enm}

In essence, there are two different types of ENMs, which differ by their dimensionality. The Gaussian Network Model (GNM), proposed by Ivet Bahar, Burak Erman and Turkan Haliloglu in 1997~\cite{Bahar:97,Erman:97}, is a one-dimension model while Tirion's model, later called the Anisotropic Network Model~\cite{Bahar:01} (ANM), is a tridimensional one. 

\subsubsection{The Anisotropic Network Model}

Although eq.~\ref{quad} may look simple, it relies on a large number of parameters, namely, the elements of the Hessian matrix (eq.~\ref{secder}).
In order to make it even simpler, M. Tirion proposed to replace eq.~\ref{quad} by another quadratic form, namely:
\begin{equation}
  \label{eqenm}
  V=\frac{1}{2} k_{enm} \sum_{d_{ij}^0 < R_c} (d_{ij}-d_{ij}^0)^2
\end{equation}
where $d_{ij}$ is the actual distance between atoms $i$ and $j$, $d_{ij}^0$ being their
distance in the studied structure~\cite{Tirion:96}. This amounts to set Hookean springs between all pairs
of atoms less than $R_c$ {\AA}ngstr\"oms  away from each other. Note that in Tirion's work, as well as in
most ANM studies (there are notable exceptions~\cite{Kneller:99}), $k_{enm}$, the spring force
constant, is the same for all atom pairs. When it is so, the role of $k_{enm}$ is just 
to specify which system of units is used,  $R_c$ being the only physically relevant parameter of the model.
In other words, when studying an ENM, the major drawback added with respect to standard NMA is that
most atomic details are simply ignored. 

However, considering eq.~\ref{eqenm} instead of eq.~\ref{quad} has several practical advantages.
First, an energy minimization is not required any more, since the configuration whose energy is the absolute minimum one
($V=0$) is known: it is the studied one. As a corollary, results obtained by studying ENMs are 
easier to reproduce. Indeed, an energy minimization not only introduces unwanted distortions in a structure,
but it does it in a way that strongly depends upon the most tiny details of the protocole used, this, also as a consequence
of the huge number of minima of a realistic PES for a biological macromolecule.    
Last but not least, as a straightforward consequence of eq.~\ref{eqenm}, the elements of the Hessian matrix (see eq.~\ref{secder}) are as simple as):
\begin{equation}
\label{eq:aij}
h_{ij} = - k_{enm} \frac{(x_i - x_j)(y_i - y_j)}{\sqrt{m_i m_j} d_{ij}^2}
\end{equation}
where $h_{ij}$ is the element corresponding to the $x$ and $y$ coordinates of atoms $i$ and $j$.

\subsubsection{The Gaussian Network Model}
\label{sec:gnm}

Because $R_c$, the cutoff value of an ANM, is usually rather small (see section~\ref{sec:cutoff}),
the corresponding Hessian matrix is sparse, that is, most of its elements (eq.~\ref{eq:aij}) are zeroes. So, as proposed by 
I. Bahar, B. Erman and T. Haliloglu~\cite{Bahar:97}, it is tempting to go another step further into
the simplification process and to consider the corresponding adjacency matrix, that is, the matrix whose elements
are:
\begin{equation}
h_{ij} = - k_{enm}
\end{equation}
when residues $i$ and $j$ are interacting ($h_{ij} = 0$ otherwise). Note that in the case of an adjacency matrix,
as well as for the Hessian matrix of an ANM, $h_{ii}$, the diagonal element $i$, is so that:
\begin{equation}
h_{ii} = - \sum_{i \ne j}  h_{ij}
\end{equation}

\begin{svgraybox}
Of course, with an adjacency matrix, information about directionality is missing. This is a major drawback of GNMs
since this means that studying a GNM can only provide informations about motion amplitudes. 
\end{svgraybox}

Note that GNMs are usually, if not always, set up at the residue level, while ANMs are
sometimes studied at the atomic level, like in the seminal study of M. Tirion~\cite{Tirion:96}.
From now on, to underline such (not so common) cases, these latter models will be coined "all-atom ANMs".

\subsubsection{The cutoff issue}
\label{sec:cutoff}

The main, if not the only, parameter of an ENM is $R_c$. Although several studies have tried 
to justify the choice of a particular value for this parameter, typically by comparing calculated and experimental quantities,
cutoff values over a wide range are still of common use, varying between 7~\cite{Phillips:02} and 16{\AA}~\cite{Phillips:07}. 

For the most part, this probably reflects the
fact the lowest-frequency modes of an ENM are usually "robust"~\cite{Nicolay:06}, that is, little sensitive to the way the model is built.
However, it is obvious that to be meaningful the
value of $R_c$ has to be on the small side. Putting it to an extreme: in the case of a GNM (see section~\ref{sec:gnm}), if $R_c$ is so large that the adjacency matrix
is completely filled with non-zero elements, its eigenvalues and eigenvectors, apart from being degenerate, will only depend upon $N$, the size
of the system, and not upon its topology or its shape. As a consequence, they can for sure not provide any useful information. On the other
hand, if $R_c$ is too small, then the network of interacting residues is split into sub-networks, either free to rotate with respect to another one (in the case of an ANM)
or completely independant from each other (in both ANM and GNM cases). Such dynamical properties are certainly not among those expected for a macromolecule,
and this is why, in ANM studies, the smallest cutoff values used are of the order of 8-10{\AA}~\cite{Tama:01,Delarue:02}, that is, larger than the
typical distance between two interacting amino-acid residues in a protein, namely 6-7{\AA}~\cite{Miyazawa:85,Miyazawa:96}.

\begin{svgraybox}
In practice, choosing a too small value for $R_c$ yields additional zero eigenvalues. 
\end{svgraybox}

So, if more than one (for a GNM) or six (for an ANM)
zero eigenvalues are obtained, then it is highly recommanded to increase $R_c$. 
Note that GNMs allow for the use of smaller values of $R_c$ (a value of 7.3{\AA} is often chosen~\cite{Phillips:02}) 
since in the case of a mono-dimensional model a single connection is enough for avoiding any free translation of a group of atoms with respect to another.
As a consequence, when a GNM is built with C$_{\alpha}$  atoms picked from a single protein chain, that is, when all amino-acid residues are chemically bonded 
to each other through peptidic bonds, a value of $R_c$ as low as 4{\AA} (the typical distance between two consecutive C$_{\alpha}$  atoms) can be used.

At first sight, it may seem that problems with small cutoff values could be solved with a distance-dependant spring force constant, as early
proposed by Konrad Hinsen~\cite{Hinsen:98}.
However, it is clear that an exponential term, for instance, introduces a typical length which, when too small, yields similar artifacts.  
Indeed, in such a case, the additional free rigid-body motions obtained with a too small value for $R_c$ are expected to be replaced by low-frequency 
motions involving the same too little-connected groups of atoms. 

Note that with ENMs other kinds of spurious low-frequency motions can be observed. For instance,
in crystal structures, protein N- and C-terminal ends are often found to extend away from the rest of the structure. As a consequence, large amplitude, usually
meaningless, motions of these (almost) free ends can be found among the lowest-frequency modes. So, in order to obtain
significant and clear-cut results, it is highly recommanded to begin an ENM study by "cleaning" the studied structure, namely, by removing such free ends.

A similar kind of spurious low-frequency motion can be observed with all-atom ANMs, in which groups of little-connected atoms are involved,
typically those at the end of long sidechains~\cite{Tama:Phd}. Note that an elegant way to cure such artifacts is to use the RTB approximation~\cite{Durand:94,Tama:00},
which allows to remove from the Hessian matrix all contributions associated to motions occuring inside each "block" the system is split into
(RTB stands for Rotation-Translation of Blocks). In most cases, a block corresponds to a given amino-acid residue but, while atom-atom
interactions are taken into account when the atoms belong to different blocks, each block can also correspond to a whole protein subunit, allowing for the
study of systems as large as entire virus capsids~\cite{Tama:02v}.

\section{Empirical foundations}
\label{sec:emp}

As illustrated above, ENMs and NMA are closely related. As a consequence, the theoretical foundations of ENMs are for the most part those of NMA.
However, when applied to complex molecular systems, NMA is known to have obvious drawbacks (see section~\ref{ssec:nma}). So,
if NMA is still widely performed it is because of its empirical, sometimes unexpected, successes. As recalled below, most of these successes
can also be achieved by studying ENMs.

\subsection{B-factors}

From eq.~\ref{nma:solution} and eq.~\ref{qampl}, it is straightforward to show that $<\Delta r_i^2>$, the
fluctuation of coordinate $i$ with respect to its equilibrium value, is so that:
\begin{equation}
\label{fluct}
<\Delta r_i^2> = \frac{k_B T}{m_i} \sum_{k=1}^{n_{nz}} \frac{a_{ik}^2}{4 \pi^2 \nu_k^2}
\end{equation}
$n_{nz}$ being the number of non-zero frequency normal modes of the system, namely,
$n_{nz}=N-1$ when a GNM is considered and $n_{nz}=3N-6$ when it is an ANM. However, in practice,
since such fluctuations scale as the inverse of $\nu_k$, the $k^{th}$ mode frequency, a sum over 
the lowest-frequency normal modes of the system is usually enough for obtaining a fair approximation~\cite{Levy:82}. 

On the other hand, $B_i$, the crystallographic Debye-Waller factor (the so-called isotropic B-factors) of atom $i$,
is expected to be related to the fluctuations of its atomic coordinates through: 
\begin{equation}
\label{eqbf}
B_i = \frac{8 \pi^2}{3} <\Delta x_i^2 + \Delta y_i^2 + \Delta z_i^2>
\end{equation}
Although other physical factors are involved, like crystal disorder or lattice phonons, as well as non-physical ones,
like the number of water molecules included in the structure refinment process by crystallographers,
significant correlations between B-factor values predicted using eq.~\ref{fluct}-\ref{eqbf} and experimentally obtained ones
have been reported in numerous cases.    

\vskip 0.5 cm
For instance, in a study of 30 protein GNMs ($R_c=7.5${\AA}), a mean value of $0.62 \pm 0.13$ for this correlation coefficient was found~\cite{Maritan:02}.
Interestingly, in the same study, 26 other proteins were considered, for which accurate relaxation measurements had been measured by NMR,
and the mean correlation between the corresponding fluctuations and those obtained using eq.~\ref{fluct} was found to be significantly higher, namely,
$0.76 \pm 0.04$,
a remarkable agreement with the experimental data being achieved in several cases, with a correlation coefficient over 0.9 for four of them~\cite{Maritan:02}. 
Amazingly,  ANMs do not perform significantly better. For instance, in a study of 83 proteins ($R_c=16${\AA}), a mean value for the correlation coefficient of $0.68 \pm 0.11$ between
predicted and isotropic B-factors was obtained~\cite{Phillips:07} while, using the all-atom ANM ($R_c=5${\AA}) implemented in the Eln\'emo webserver~\cite{Elnemo2},
which makes use of the RTB approximation~\cite{Durand:94,Tama:00}, a very similar value of  $0.68 \pm 0.13$ was found~\cite{Phillips:07}. 

\vskip 0.5 cm
Note that in both studies mentioned above, when eq.~\ref{fluct} was used, overall translations or rotations of the entire protein within the crystal cell were excluded from
the calculation, while it is well known that such motions are able to provide by themselves good correlations with experimental values~\cite{Kuriyan:91}. 
In other words, much better correlations with experimental B-factors can be obtained by mixing NMA predictions with protein rigid-body motions,
the latters accounting partly for crystal disorder, but mostly for the phonon modes of the whole crystal. Interestingly, these latter modes can be taken into
account within the frame of ENM studies, simply by including all crystal cell symmetries in the model~\cite{Simonson:92,Hinsen:08,Phillips:09}.

Of course,  such significant correlations with experimental data can only be obtained because the amplitude of atomic thermal fluctuations scales
as the inverse of mode frequencies (see eq.~\ref{fluct}). Indeed, with crude models like ENMs, the actual high-frequency modes of a protein can not be predicted,
 because such modes strongly depend upon the chemical details of the structure, only a few neighboring atoms (e.g., covalently bonded ones) being involved
in the highest-frequency modes.  This does not mean, though, that the high-frequency modes of an ENM can not bring any useful information.
Indeed, they correspond to local motions occurring within the parts of the structure whose density is the highest~\cite{Erman:97}. Moreover, it has been shown that 
such regions often ly nearby enzyme active sites~\cite{Juanico:07,Lavery:07}.

On the other hand, the B-factor values themselves can not directly be obtained by studying ENMs, since their average is proportional to $k_{enm}$.
Indeed, it is customary to choose $k_{enm}$ so as to match average experimental B-factor values~\cite{NMA}. Another common way is to choose $k_{enm}$ so as to reproduce
the lowest-frequency of the system, as obtained using all-atom force-fields~\cite{Juanico:07}.

\begin{figure}[t]
\includegraphics[scale=.65]{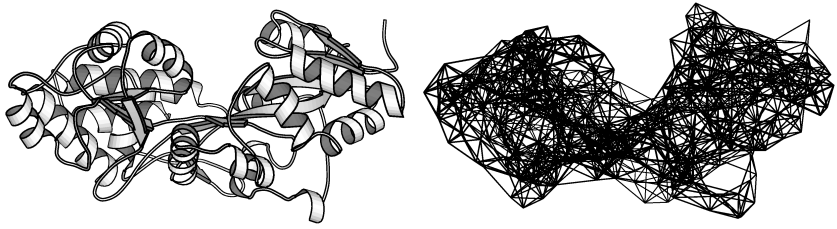}
\caption{Left: the open (ligand-free) form of maltodextrin binding protein (PDB identifier 1OMP). Right: the corresponding Elastic Network Model. Pairs of C$_{\alpha}$ atoms
are linked by springs (plain lines) when they are less than 8{\AA} from each other. Drawn with Molscript~\cite{Molscript}.}
\label{fig:mbp}     
\end{figure}

\subsection{The relationship with protein functional motions}

The seminal paper of M. Tirion ends with the statement that~\cite{Tirion:96}:
\begin{quotation}
Tests performed on a periplasmic maltodextrin binding 
protein (MBP) indicate that the slowest modes do indeed closely map the open
form into the closed form (Tirion, in preparation).
\end{quotation}
The next paper of M. Tirion never came out but her result was confirmed a few years later,
as part of a study of 20 protein ENMs ($R_c=8${\AA}) in both their ligand-free (open) and ligand-bound (closed) forms~\cite{Tama:01}. Indeed, for MBP, it was found that the
overlap between its second lowest-frequency mode and its functional conformational change is close to 0.9. 
This means that 80\% of the functional motion
of MBP can be described by varying the normal coordinate associated to a single of its modes. Indeed, $O_k$, the overlap with mode $k$, is given by:      
\begin{eqnarray}
\label{overlap}
O_k = 
\frac {\sum_i \Delta r_i a_{ik}}
{\sqrt{\sum \Delta r_i^2}}
\end{eqnarray}
where $\Delta r_i$ is the variation of coordinate $i$ between the open and the closed form
after both structures have been superimposed~\cite{Marques:95}. On the other hand, since the modes of MBP form
an orthogonal basis set, the following property holds:
\begin{eqnarray}
\sum_{k=1}^{n_{nz}} O_k^2 = 1 
\end{eqnarray}

More generally, it was found that when the conformational change of a protein upon ligand binding happens to be highly collective, one of its low-frequency normal modes
often compares well with the experimental motion (overlap over 0.5~\cite{Tama:01}). 
Since then, a study of nearly 4,000 cases has confirmed this result~\cite{Gerstein:02}, while another study of a set of proteins with similar functions and shapes, but various
folds, namely DNA-dependant polymerases~\cite{Delarue:02}, has shown that the low-frequency modes of a protein, and hence the nature of its 
large amplitude motions, are likely to be determined by its shape~\cite{Tama:01,Ma:05b,Tama:06}. 

\vskip 0.5 cm
Indeed, this latter point has recently been confirmed in a rather direct way, by considering ENMs built in such a way that each amino-acid interacts with a given
number of neighbors (the closest ones). Then, at variance with cutoff-based ENMs, the rigidity of the system is fairly constant from a site to another.
However,  the relationship between the lowest-frequency modes of a protein and its functional motion is preserved. Specifically,
it was found that the subspace defined by up to the 10-12 lowest-frequency modes of a protein is conserved, whatever model is used.
Moreover, when no such, so-called robust, subspace exists, the fonctional motion of the protein is found to be either localized and/or of
small amplitude (typically: less than 2-3{\AA} of C$_{\alpha}$-r.m.s.d)~\cite{Nicolay:06}.

\vskip 0.5 cm
In retrospect, these results make sense. First, a strong relationship between low-frequency modes and protein fonctional motions
was first observed within the frame of NMA studies performed at a highly detailed, atomic level of description, noteworthy in the cases
of lysozyme~\cite{Karplus:76}, hexokinase~\cite{Harrison:84}, citrate synthase~\cite{Marques:95} and hemoglobin~\cite{Perahia:95}.
Since, as recalled above, it was later found that such a relationship also holds when most chemical details are removed, it is clear that the property captured by NMA has to be
a very general one. On the other hand, K. Hinsen has convincingly shown that the low-frequency modes of a protein can be used 
to split its structure into well-defined domains~\cite{Hinsen:98}, with the additional advantage of a smooth, almost continuous, description of their boundaries. 
So, since it is well known that most large amplitude protein functional motions
can be well described as combinations of almost rigid-body motions of entire structural domains~\cite{Branden:91,Gerstein:98}, the relationship
found between these motions and the low-frequency modes of ENMs is just another demonstration that whole quasi-rigid domain motions 
are involved in such modes. On the other hand, it is not that difficult to admit that the spatial clustering of amino-acids into domains 
can be revealed by studying protein dynamical properties, even at a crude level of description. A corollary of this line of thought is that
ENMs should perform better, as far as low-frequency and large amplitude motions are concerned, in the case of large, multi-domain systems.   

\subsection{Applications}

As illustrated above, NMA of ENMs seems to have a clear predictive power. 
So, given both the simplicity of these models and their coarse-grained nature, many applications have been proposed.
For instance, as early suggested, being able to guess the pattern of atomic fluctuations through eq.~\ref{fluct}
may prove useful for refining crystal structures~\cite{Diamond:90,Kidera:90}. 

However, most applications take advantage of the possibility to predict atomic displacements through the reciproqual of eq.~\ref{qr}, namely:
\begin{equation}
\label{rq}
r_i = r_i^0 + \frac{1}{\sqrt{m_i}} \sum_k^{n_{sub}} a_{ik} q_k
\end{equation}
where $n_{sub}$ is the number of low-frequency modes considered to be enough for performing an accurate prediction.
In the simplest case, mode amplitudes can be varied arbitrarily, one mode after the other. Indeed, in the light of enough experimental data, the analysis of such trajectories
can prove enough for getting insights about the nature of the functional motion of a protein~\cite{Helene:03,Sanejouand:96}. 
Some of the conformations thus obtained can also allow for solving difficult molecular-replacement problems, although
it is often necessary to explore at least a couple of modes in order to reach a useful conformation~\cite{Elnemo1}. More generally,
eq.~\ref{rq} can be used so as to reduce the dimensionality of the system and, thus, to find more easily protein conformations
fulfiling a given set of constraints. For instance, it has been used for fitting known structures into low-resolution electron density maps~\cite{Delarue:04,Tama:04,Hinsen:05, Norma}
providing, for instance, more detailled structural data for systems of major interest, like the ribosome~\cite{Tama:05}.  

\begin{svgraybox}
Note that eq.~\ref{rq} is linear. As a consequence, atom motions follow straight lines and local distorsions (of most chemical bonds, valence angles, etc) can
not be avoided. So, for many applications, as well as for obtaining well-behaved normal mode trajectories, the conformations thus
generated need to be "regularized"~\cite{Elnemo1}, using for instance a detailled all-atom force-field and standard energy-minimization techniques. 
\end{svgraybox}

\section{Conclusion}

Fifteen years after their introduction in the realm of molecular biophysics~\cite{Tirion:96}, thanks to their simplicity as well as to their coarse-grained nature, Elastic Network Models are 
becoming more and more popular. Indeed, many applications have been proposed, noteworthy within the frame of various structural biology techniques. 

From a theoretical point
of view, their relationship with Normal Mode Analysis is obvious, since both approaches rely on a quadratic form for the energy function, the former, {\it par d\'efinition}, 
the latter, as a consequence of a small displacement, so-called harmonic (or linear) approximation. 

From an empirical point of view, it has been extensively shown
that normal mode studies of Elastic Network Models yield low-frequency, large amplitude and collective, motions which prove often similar 
to those obtained with an all-atom model and a standard empirical force-field. 

This is likely to be a consequence of the robusteness of these motions~\cite{Nicolay:06}. 
Moreover, such motions
often provide fair predictions for the pattern of thermal atomic fluctuations (e.g. the crystallographic B-factors) or for the kind of functional motion a given protein can perform (e.g.
its conformational change upon ligand binding).  



\end{document}